\documentclass[twocolumn]{aastex63}

\usepackage{fontawesome}
\usepackage{CJK}
\usepackage{lipsum}
\usepackage{xcolor}

\usepackage{graphicx}
\begin{document}
\begin{CJK*}{UTF8}{gbsn}
\title{Fizzy Super-Earths: Impacts of Magma Composition on the Bulk Density and Structure of Lava Worlds}

\email{boley.62@osu.edu}

\author[0000-0001-8153-639X]{Kiersten M. Boley}
\altaffiliation{NSF Graduate Research Fellow}
\affiliation{Department of Astronomy, The Ohio State University, Columbus, OH 43210, USA}

\author[0000-0001-5753-2532]{Wendy R. Panero}
\affiliation{School of Earth Sciences, The Ohio State University, Columbus, OH 43210, USA}

\author[0000-0001-8991-3110]{Cayman T. Unterborn}
\affiliation{Space Science \& Engineering, Southwest Research Institute, San Antonio, Texas}

\author[0000-0003-3570-422X]{Joseph G. Schulze}
\affiliation{School of Earth Sciences, The Ohio State University, Columbus, OH 43210, USA}

\author[0000-0003-1445-9923]{Romy Rodr\'iguez Mart\'inez}
\affiliation{Department of Astronomy, The Ohio State University, Columbus, OH 43210, USA}

\author[0000-0002-4361-8885]{Ji Wang (王吉)}
\affiliation{Department of Astronomy, The Ohio State University, Columbus, OH 43210, USA}




\begin{abstract}
Lava worlds are a potential emerging population of Super-Earths that are on close-in orbits around their host stars with likely partially molten mantles.  To date, few studies address the impact of magma on the observed properties of a planet. At ambient conditions magma is less dense than solid rock; however, it is also more compressible with increasing pressure. Therefore, it is unclear how large-scale magma oceans affect planet observables, such as bulk density.  We update \texttt{ExoPlex}, a thermodynamically self-consistent planet interior software, to include anhydrous, hydrous (2.2 wt\% H$_2$O), and carbonated magmas (5.2 wt\% CO$_2$).  We  find that Earth-like planets with magma oceans larger than $\sim$ 1.5 R$_\oplus$ and $\sim$ 3.2 M$_\oplus$ are modestly  denser than an equivalent mass solid planet. 
From our model, three classes of mantle structures emerge for magma ocean planets: (1) mantle magma ocean, (2) surface magma ocean, and  (3) one consisting of a surface magma ocean, solid rock layer, and a basal magma ocean. The class of planets in which a basal magma ocean is present may sequester dissolved volatiles on billion-year timescales, in which a 4 $M_\oplus$ mass planet can trap more than 130 times the mass of water than in Earth's present-day oceans and 1000 times the carbon in the Earth's surface and crust.

\end{abstract}
\section{Introduction}
As terrestrial planets evolve, they are thought to transition through a magma ocean phase directly following formation \citep{Schaefer2018,deVries2016,Elkins-Tanton2012a}. During this phase, the temperatures are sufficiently high for global melting of the silicate mantle \citep{Chao2021, Elkins-Tanton2012a}. Although global magma oceans would generally be considered a transient phase in the lifetime of a terrestrial planet, it is possible for them to be long-lived. Particularly, planets on close-in orbits will experience longer magma ocean lifetimes maintaining their surface temperatures through the radiation of their host star \citep{Chao2021}. Ultra short period (USP) planets orbiting a host star with a strong magnetic field are also subject to an additional form of heating due to induction that could produce magma layers in the upper mantle as well \citep{Kislyakova2020,Kislyakova2018,Kislyakova2017}. These non-transient magma oceans or lava worlds undergo a distinct evolution that is unique to their temperatures and orbital distance.

After formation, many of these planets will initially experience significant atmospheric mass loss due to the high irradiation of their host stars making them unlikely to maintain a substantial atmosphere \citep{Lopez2012}. There are several works that find USP super-Earths that are consistent with the absence of such atmospheres \citep{Keles2022,Kreidberg2019}. Although these planets would require rare initial conditions to maintain their primary atmosphere so close to their star, some planets may still have partial atmospheres due to the vaporization of silicate rock on their permanent day-side as a result of tidal locking \citep{Zieba2022, Leger2011}. During their evolution, they inevitably become tidally locked with their host star resulting in a day- and night-side \citep{Rory2017,Kasting1993}. There is evidence that the day-side may possess an atmosphere at low pressures (P $\leq$ $10^{-5}$ bar,1.5 Pa) consisting of rock vapor \citep{Zieba2022, Leger2011}. The proximity to their host star makes short-period planets not only more easily detectable but also optimal for characterization.

The vast majority of likely terrestrial planets with masses $\lesssim$ 10 M$_\oplus$ or radii $\lesssim$ 1.7 R$_\oplus$ discovered are on orbits less than 10 days. These discoveries have naturally prompted the characterization of these planets to understand their properties.  For the majority of terrestrial planets without atmospheres, the stellar abundances of the host stars are correlated with the bulk density of the planet \citep{Schulze2021, Adibekyan2021}. Previous studies specifically focus on solid planets while including the low densities with high equilibrium temperatures of some of the planets in their samples, e.g., 55 Cnc e and WASP-47 \citep{Schulze2021,Adibekyan2021,Becker2015,McArthur2004}. In these works, high-density planets are explained by an excess of iron. However, the description of low-density planets is largely unexplored due to their non-unique relationships between planet composition and density. 


There are several reasons in which one might attempt to explain the low-density of these planets using magma oceans. The first of which is that the surface temperatures of many of these planets are above 1350 K approximately the zero-pressure melting temperature of Earth-like rocks \citep{Katz2003}. Therefore, their surfaces may exceed the melting temperature of typical rock, which range from 900 K-1500 K making them unlike any planet in our solar system \citep{Liu2010, Hirose1993, Takahashi1983}.
Materials also expand when they are heated and melt; however, magmas are also highly compressible potentially counteracting any expansion at depth \citep{McCormick2016,Rivalta2007}. While Earth-like magmas are up to 24 \% less dense than their solid at the surface, they experience 20\% compression compared to their zero-pressure volume at a pressure equivalent to 150 km below the Earth's surface (5 GPa)\citep{Dannberg2016}, compared to a 6\% compression of a solid of the same composition. As a result, it is possible that planets with deep magma ocean are denser than an equivalent mass solid planet. However, it is not well understood the degree to which this would impact the bulk-density of the planet, and whether this difference is observable.

{This paper will demonstrate that global magma oceans have little impact on bulk density of likely lava worlds. We will quantify the degree to which melting can lower the observed densities.} For close-in Super-Earths with masses $\gtrsim$ 6 M$_\oplus$, such as 55 Cnc e \citep{McAuthor2004} and  WASP-47 e \citep{Vanderburg2017}, there have been many studies that propose several explanations for their low densities which range from the presence of an atmosphere to a core-free planet \citep{Dorn2021,Dorn2019, Demory2016, ito2015,Gillion2012}. However, there has been little investigation on the impact of magma on planet density, independent of an atmosphere. This is particularly important for less massive planets that are too small to maintain a substantial atmosphere, such as TOI-561 b \citep{Lacedelli2021}.

{We will address the impact of magma on the mantle structure of a lava world. Given that volatile species readily dissolve within magma, we explore the implications for trapping these volatiles in the interior over planetary lifetimes with implications for outgassing. Previous studies have shown evidence for basal magma oceans to occur in likely lava worlds and during the phases of Earth's evolution \citep{Labrosse2007, ANDRAULT2019}. H$_2$O and CO$_2$ are volatiles thought to be commonly outgassed from the mantle during the thermal evolution of a planet to form the early atmosphere \citep{Schaefer2017, Lupu2014}. By dissolving H$_2$O or CO$_2$ within magma, we can explore the amount of volatiles that may be stored on billion-year timescales. Therefore, we consider the mass-surface temperature relationship exploring the mantle structures that arise and whether volatiles may be sequestered within the mantle.}

The outline of our paper is as follows. Section \ref{s:method} provides a description of our approach and additions to \texttt{ExoPlex} 
\footnote{\url{ https://github.com/CaymanUnterborn/ExoPlex/}} \citep{2018NatAs...2..297U, Unterborn2019, Unterborn2023}. In Section \ref{s:MR}, we discuss the resulting density-radius relationships comparing our findings to likely magma ocean planets, and the impact of magma composition on the bulk density and structure of a planet.  We discuss the limitations of our model in Section \ref{s:Discuss} along with comparing our results to previous studies and known exoplanets. Finally, we provide a summary and conclude in Section \ref{s:summary and conclusions}.

\section{Method} \label{s:method}

The base of our planet interior model uses \texttt{ExoPlex}, a thermodynamically self-consistent planet interior software \citep{2018NatAs...2..297U, Unterborn2019, Unterborn2023}. \texttt{ExoPlex} solves five coupled differential equations: the mass within a sphere, hydrostatic equilibrium, adiabatic temperature profile, Gauss’ law of gravity in one dimension, and the thermally-dependent equation of state for solid planets. We construct a magma module within this software to include the liquid phase of the mantle when the temperature exceeds the melting point. As melting temperatures are compositionally dependent, we constrain this module to Earth-like mantle compositions. In this study, we assume a spherically symmetric planet for simplicity. However, tidal locking would cause lava worlds to have drastic temperature differences between the substellar and anti-stellar point \citep{Leger2011}. Therefore, this module considers the upper limit that magma may have on a planet's observed properties. 


\subsection{Mantle}

We adopt an Earth-like pyrolitic molar composition of 0.5Na$_2$O- 2CaO- 1.5Al$_2$O$_3$- 4FeO -30MgO -24SiO$_2$, which is within 1 wt $\%$ of the bulk silicate composition of the Earth \citep{Solomatova2020, MCDONOUGH1995}. For a more realistic model, we incorporate both the solid and melt phases of pyrolite. To simplify, we neglect chemical fractionation assuming the melt phase to have a liquid pyrolite composition at all temperatures above the melt curve. We also include the effects of volatiles within the melt phase using three separate model scenarios: Anhydrous, Hydrous, and Carbonated models. 

In the multi-component systems of planetary mantles, melting occurs over a range of temperatures. The solidus is the temperature at which the melting begins with a very small volume fraction of melt. As the temperature increases, the volume fraction of melt also increases until all components of the rock are liquid. The temperature at which the rock becomes completely liquid is referred to as the liquidus. For most silicate rocks, the solidus is hundreds of degrees lower than the liquidus. When a rock undergoes a small degree of melting just above the solidus, the trapped volatile species tend to enter the melt phase \citep{Hirschmann2000}. As one of the objectives of this study is to determine if a magma ocean is observable via the bulk density of a planet, use of the solidus as the temperature at which the planet melts will overestimate the density effects of melting. However, it is solid layers, forming a mechanical barrier, that will sequester volatiles deep in the interior. Therefore, we choose to focus on the solidus temperature acknowledging that the effects on density are lower bounds.

Many previous studies that incorporate the effects of melt in exoplanets have used single-component melt curves (e.g. \cite{Dorn2021,Dorn2019,Stixrude2014}). However, multi-component systems are more likely to occur given the diversity of rock-forming refractory materials available within the protoplanetary disk during formation \citep{Helling2014}. Therefore, we adopt the solidus temperature at which significant changes in physical properties occur. Specifically, we model the solidus similar to the method used in \cite{Boukare2022}. \cite{Boukare2022} rely on solidus and liquidus curves from \cite{Zhang1994} and \cite{Fiquet2010}. Adopting a similar method, we use a solidus that is lower than \cite{Zhang1994} using methods that overcome experimental limitations of detecting small volumes of melts \citep{Nomura2014}.

\begin{figure}[!t]
\begin{center}
\includegraphics[width=\linewidth]{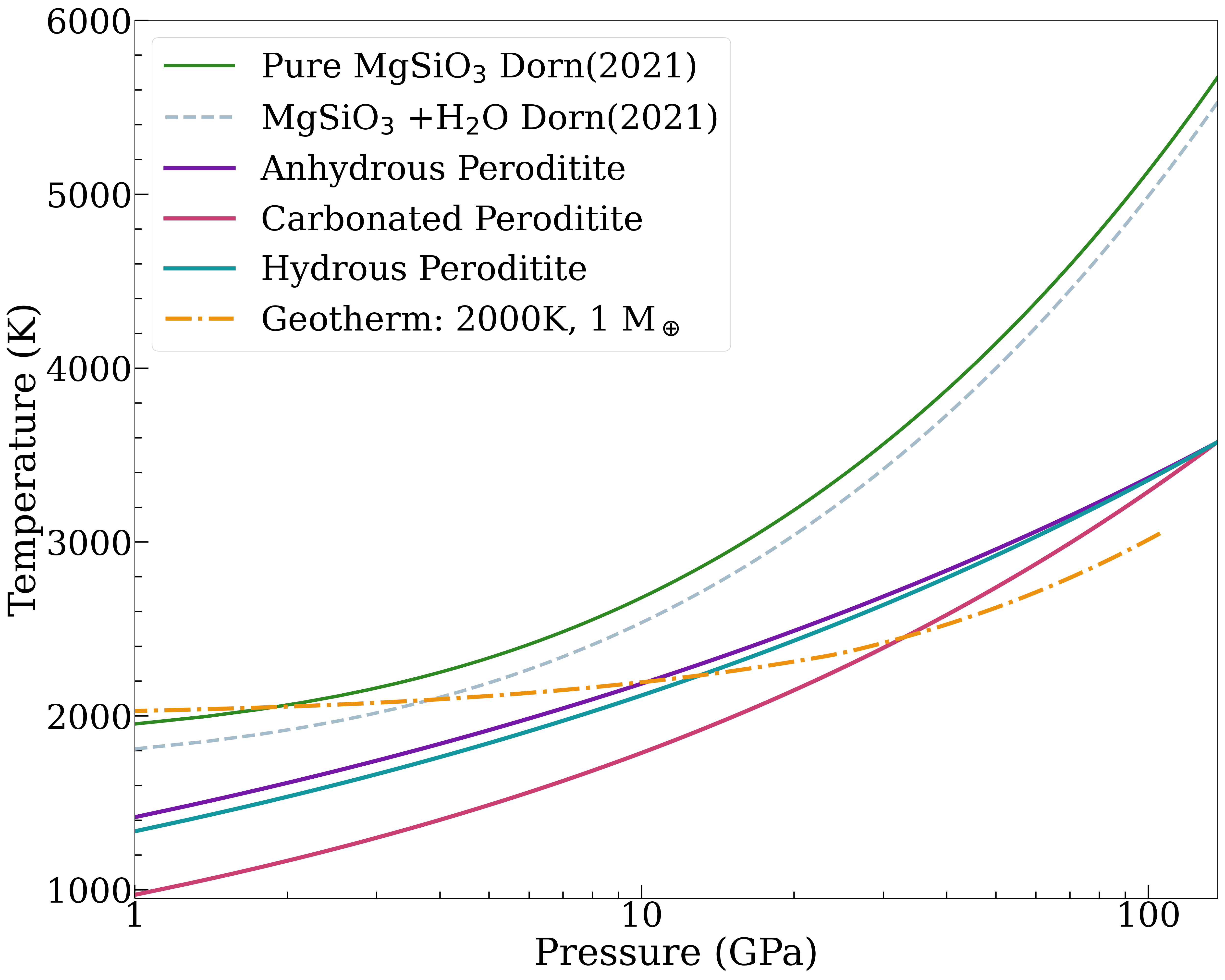}
\caption{Solidus melt curves for the three magma compositions: Anhydrous (purple), Hydrous (Cyan), Carbonated (Dark Blue). We include the pressure-temperature profile of a planet at a surface temperature, $T_{surf}=2000$ K and 1 Earth mass. These are compared to the melt curves within \cite{Dorn2021}, which uses a piece-wise solidus melt curve. }
\label{fig:Melt}
\end{center}
\end{figure}

To determine the solid-melt transition within the mantle, we use a peridotite solidus melt curve from \cite{Katz2003} for pressures less than 10 GPa for the anhydrous and hydrous compositions. This gives us the following equation for the solidus melt within the mantle: 

\begin{equation} \label{eq:Kaz}
    T_{Melt,Anhydrous}= a_1 P^2 + a_2 P+ a_3
\end{equation}
\\
where $P$ is pressure, $a_{1}=-5.1 ^{\circ}$C GPa$^{-2}$, $a_{2}=132.9 ^{\circ}$C  GPa$^{-1}$, $a_{3}=1358.85$ K. 

For the hydrous melt, we use the \cite{Katz2003} solidus melt curve for hydrous melts, which has an additional term to account for H$_2$O within the melt: 
\begin{equation}
    T_{Melt,Hydrous}= a_1 P^2 + a_2 P + a_3 - K X^\gamma_{H_2O}
\end{equation}
\\
where $K= 43 ^{\circ}$ C wt\%$^{-\gamma}$ and $\gamma=0.75$. We assume the wt\% of $H_2O$ in the melt is $X_{H_2O}=2.2$ wt\%, which corresponds to the wt\% used in \cite{Solomatova2020} hydrous magma equation of state.

To extend each solidus to higher pressures for all melt compositions, we fit a power law similar to the Simon melting law \citep{Stixrude2014,Ross1969}.

\begin{equation} \label{eq:Simon}
    T_{Melt}= T_{ref} \left( \frac{P}{P_{ref}}\right)^{b}
\end{equation}
\\
where $T_{ref}$ and $P_{ref}$ are the reference temperature and pressure at higher pressures. We adopt the anhydrous melt temperature and pressure from \cite{Nomura2014} of $T_{ref}=3570$ K at $P_{ref}=136$ GPa. We use these values   to provide an upper bound for the transition between solid and melt at higher pressures. We then fit for $b$ so that the computed melt curve intersects the peridotite solidus temperature of $T_{ref}=3570$ K at $P_{ref}=136$ GPa along with the solidus melt curves at lower pressures. We find the $b$ for the anhydrous melts to be $b_{anhydrous}=0.188$ above 10 GPa.  For the hydrous melt curve, we similarly use the \cite{Nomura2014} constraint as an upper bound for the hydrous melting between 10 and 136 GPa. As melting temperatures are closely related to density, differences in the volatile content of the melt for pyrolitic compositions become less prominent at higher pressures. Therefore, anhydrous, carbonated, and anhydrous melts converge at higher pressures. For the hydrous model above 10 GPa, we find $b_{hydrous}=0.201$.

For the carbonated melt, we use a similar approach as the above, fitting 5.2 wt\% carbonated peridotite solidus data from 0-10 GPa  \cite{Dasgupta2006} and the \cite{Nomura2014} high-pressure point (i.e $T_{ref}=3570$ K at $P_{ref}=136$) to the Simon melting law in equation (\ref{eq:Simon}) to derive $b$ for the carbonated melt to be $b_{carbonated}=0.265$.

The consequences of extrapolating these melt curves beyond \cite{Nomura2014} are minor as the difference in compression at such high pressures becomes modest, and our structural model is insensitive to the location of the melt curve at higher pressures. Our computed melt curves are shown in Figure \ref{fig:Melt}.

\begin{table}[t]
\centering
\begin{tabular}{ccccc}
\hline
Melt Composition & $\rho_0$ [$g/cm^3$]& $K_0$ [GPa] & $K_0^\prime$ & wt $\%$\\
\hline
\hline
Pyrolite & 2.49 & 24 & 4.7 & -\\
Pyrolite+ 4 $H_2O$ & 2.32 & 15 & 5.5& 2.2\\
Pyrolite+ 4 $CO_2$ & 2.37 & 18 & 5& 5.2\\
\hline
\end{tabular}
\caption{ExoPlex Model Parameters recalculated from \cite{Solomatova2020} for magma with Earth-like compositions, and with added 2.2 wt\% H$_2$O and 5.2 wt\% CO$_2$.  Each value in the table is at a reference temperature of 2000K fit using the third-order Birch-Murnagham EOS}. For all compositions, we find a best-fit thermal equation of state of $\gamma_0 =  1.7$ , $q = 0.93$, and $\theta_{D,0}=1000 K$ 
\label{tab:Params}
\end{table}

To calculate the EOS of the solid mantle, we use \texttt{ExoPlex}. \texttt{ExoPlex} relies on a fine mesh grid approach to calculate the stable mantle mineral assemblage for a given pressure and temperature from \texttt{PerpleX}~\citep{Connolly2009} grids. We construct a grid consistent with the pyrolitic composition used in \citep{Solomatova2020} as we also use their data for the melt within the mantle. Within \texttt{ExoPlex}, we set the mantle composition to reproduce the pyrolytic composition as in \cite{Solomatova2020} using the following molar fractions: Ca/Mg = 0.067, Si/Mg = 0.8, Al/Mg = 0.05, and Fe/Mg = 0.9. Given that the molar fraction of iron within \texttt{ExoPlex} includes the core and mantle, we specify a mass fraction of 0.079 wt\% FeO within the mantle to remain consistent with \cite{Solomatova2020}. 

For the melt, we use ab initio molecular dynamics data for pyrolite from \cite{Solomatova2020} shown in Table \ref{tab:Params}.  Within \texttt{ExoPlex}, we recast Solomatova and Caracas’ series of isothermal equations of state between 2000K and 5000K to a single pressure (P)-density ($\rho$)- temperature (T) equation of state model for each composition by fitting their results as follows,

\begin{equation}\label{eq:P}
    P(\rho,T)=P(\rho_0,2000 K )+ P_{th}
\end{equation}
\\
where $P(2000 K)$ is calculated by the BM3 equation and the 2000 K $\rho_0$ is as in Table \ref{tab:Params}. The thermal pressure ($P_{th}$) is expressed by the Mie-Gr{\"u}neisen relation
{
\begin{equation}
    P_{th}=\frac{\gamma}{V} \left [ E(T, \theta_D) -E(2000 K, \theta_D) \right]
\end{equation} }
\\
where $\gamma$ is the Gr{\"u}neisen parameter,$\gamma=\gamma_0 (\rho_0/\rho)^q$, and  the Debye temperature, $\theta_D=\theta_{D,0} exp[(\gamma_0 - \gamma)/q]$. The harmonic internal energy E(T,$\theta_D$) is calculated from the Debye model (e.g. \cite{Fei1992}).

{ 
To determine the best-fit the thermal parameters, we use a Markov Chain Monte Carlo (MCMC). For all compositions, we use the higher temperature calculations (up to 5000 K) to constrain $\gamma$ and $q$. We assume $\theta_{D,0}=1000$ K as fluctuation in $\theta_{D,0}$ accounts for minimal differences in $\gamma$ and $q$. Within each step of the MCMC a $\gamma$ and $q$ are proposed, we then use the \cite{Solomatova2020} 2000K data to determine the pressures and densities for the higher temperature (up to 5000 K). The densities at 135 GPa are compared to \cite{Solomatova2020} data at higher temperatures using the 2 sample Anderson-Darling statistic\citep{Anderson1952,Pettitt1976}. We find the best-fit thermal parameters for all compositions to be: $\gamma_0 =  1.7$, $q = 0.93$, and $\theta_D=1000 K$. }

Using the best-fit thermal parameters, we self-consistently calculate the interior temperature along an adiabat that corresponds to the surface temperature of the planet.  We, therefore, fit the suite of pressure (P) - density ($\rho$) - temperature (T) data to the BM3 - Mie Gr{\"u}neisen equation of state at a reference temperature of 2000 K  for dry, hydrous, and carbonated pyrolite from \cite{Solomatova2020} as shown in Table \ref{tab:Params}.  We then determine the density along the adiabat drawing from the melt or solid EOS depending on the relative position of the adiabatic temperature and the solidus. 

\begin{figure}[!t] 
\begin{center}
\includegraphics[width=\linewidth]{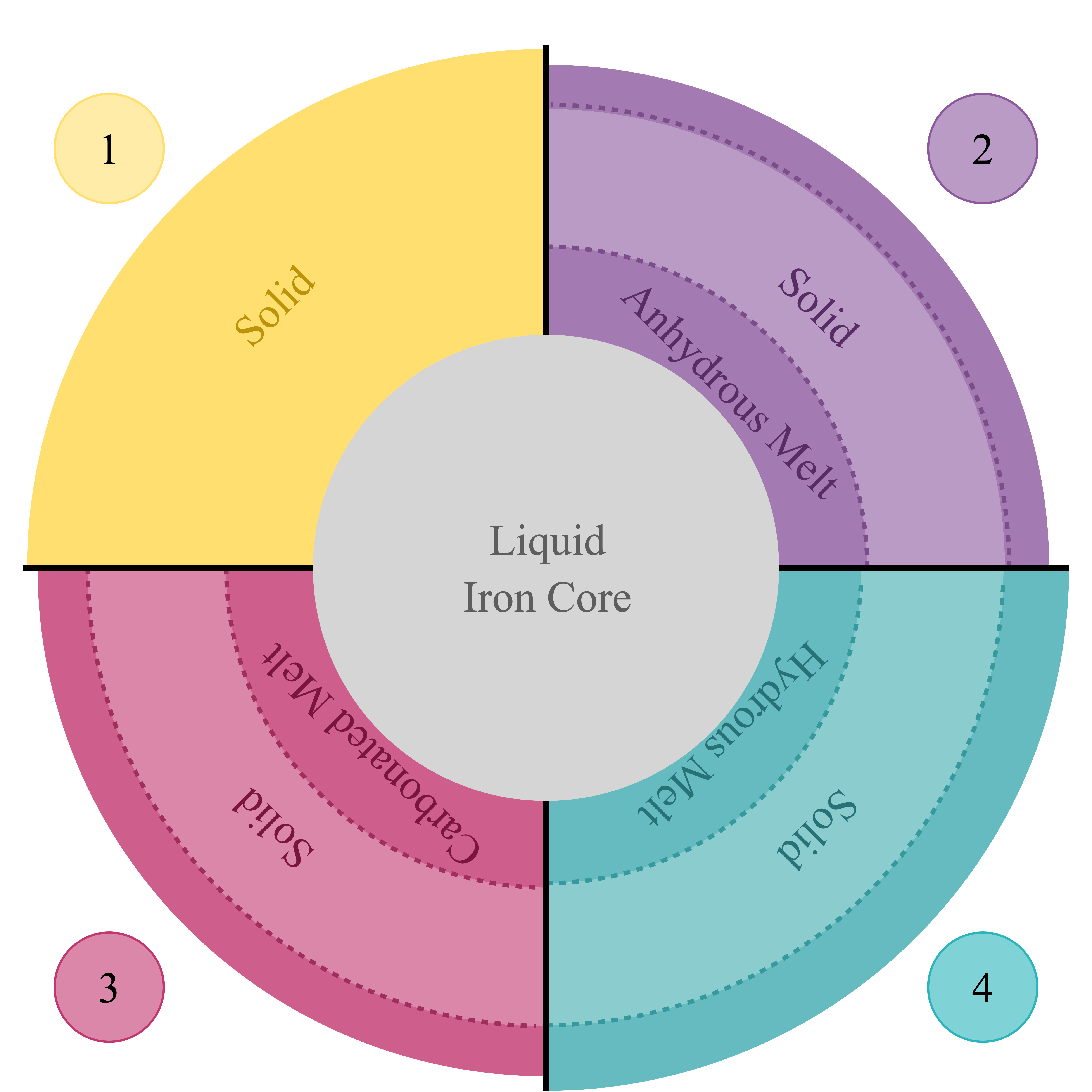}
\caption{We show an example of the four model scenarios using the results of our study of 4 M$_\oplus$ planet with a surface temperature of 2000 K. Model (1) is a solid rocky planet, where liquid rock phases are neglected. For Model (2), we introduce a molten phase within the mantle in addition to the mantle. Model (3) has an additional volatile CO$_2$ making it a carbonated melt. Model (4), the hydrous melt, generally exhibits the largest increase in radius of the four scenarios.}
\label{fig:Pslice}
\end{center}
\end{figure}

\subsection{Metallic Core}

{We explore two core scenarios as an upper and lower bound on the density variation that may arise from the core instead of mantle melting. We assume an Earth-like core composition. Given the surface temperatures required to produce a global magma ocean, our model assumes a liquid metallic core as the core-mantle temperatures will exceed that of the melting temperature of iron.

As an upper bound on the bulk density, we assume a pure liquid iron core. For this scenario, we rely on the liquid Fe pressure and temperature grid within \texttt{ExoPlex}. This grid uses the temperature-dependent equation of state of \citet{Ander94} and is calculated up to 15 TPa and 10,000 K \citep{Unterborn2023}. 

We include a lower bound on the bulk density with the addition of lighter elements in the core which results in a decrease in the bulk density of a planet. We use FeO within our models with realistic mass fractions within the core: Fe=84\% and O=16\%. We choose these values to represent a lower bound on the density as the highest estimate for light alloys in the core of Earth is $\sim$ 16\% by mass \citep{Hirose2022}. \texttt{ExoPlex} assumes that, to first-order, the presence of light elements lowers only the molar weight of liquid Fe while not affecting its compressibility at pressures indicative of exoplanetary cores. This assumption means that for an Fe core containing 16 wt\% O, the density of the planet would be  $\sim7\%$ lower than a planet with a pure Fe core for a 1 M$\oplus$ solid planet.

\begin{figure}[!t]
\begin{center}
\vspace{0.87cm}
\includegraphics[width=\linewidth]{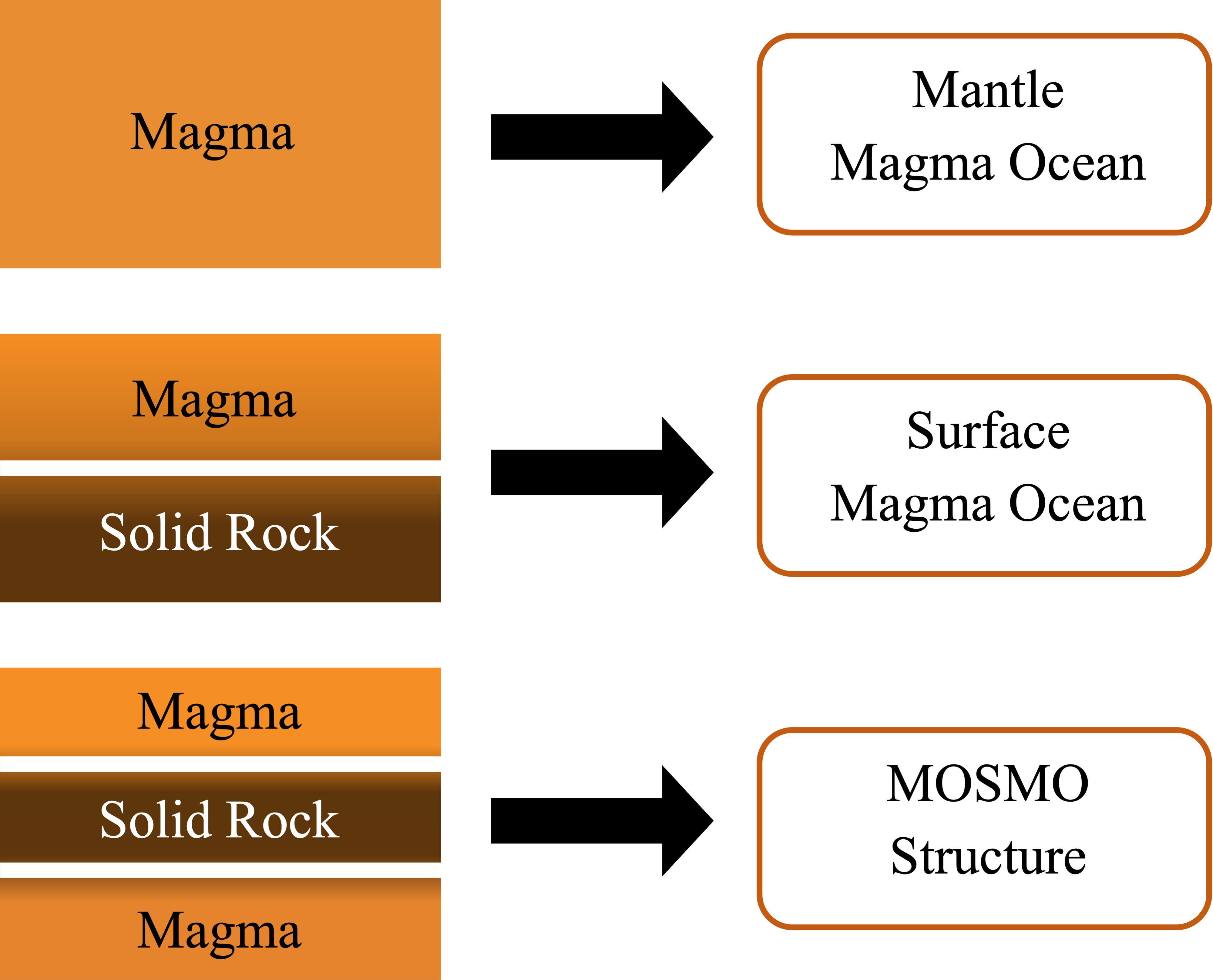}
\caption{Mantles Structures produced by our model. We find that the mantle may be mantle magma ocean, a surface magma ocean and solid rock layer, or a MOSMO structure (i.e. Surface Magma Ocean (MO)-Solid Rock Layer (S)- Basal Magma Ocean (MO)) }
\label{fig:Structure}
\end{center}
\end{figure}
\begin{figure*}[!t]
\centering
\includegraphics[width=1.1\linewidth]{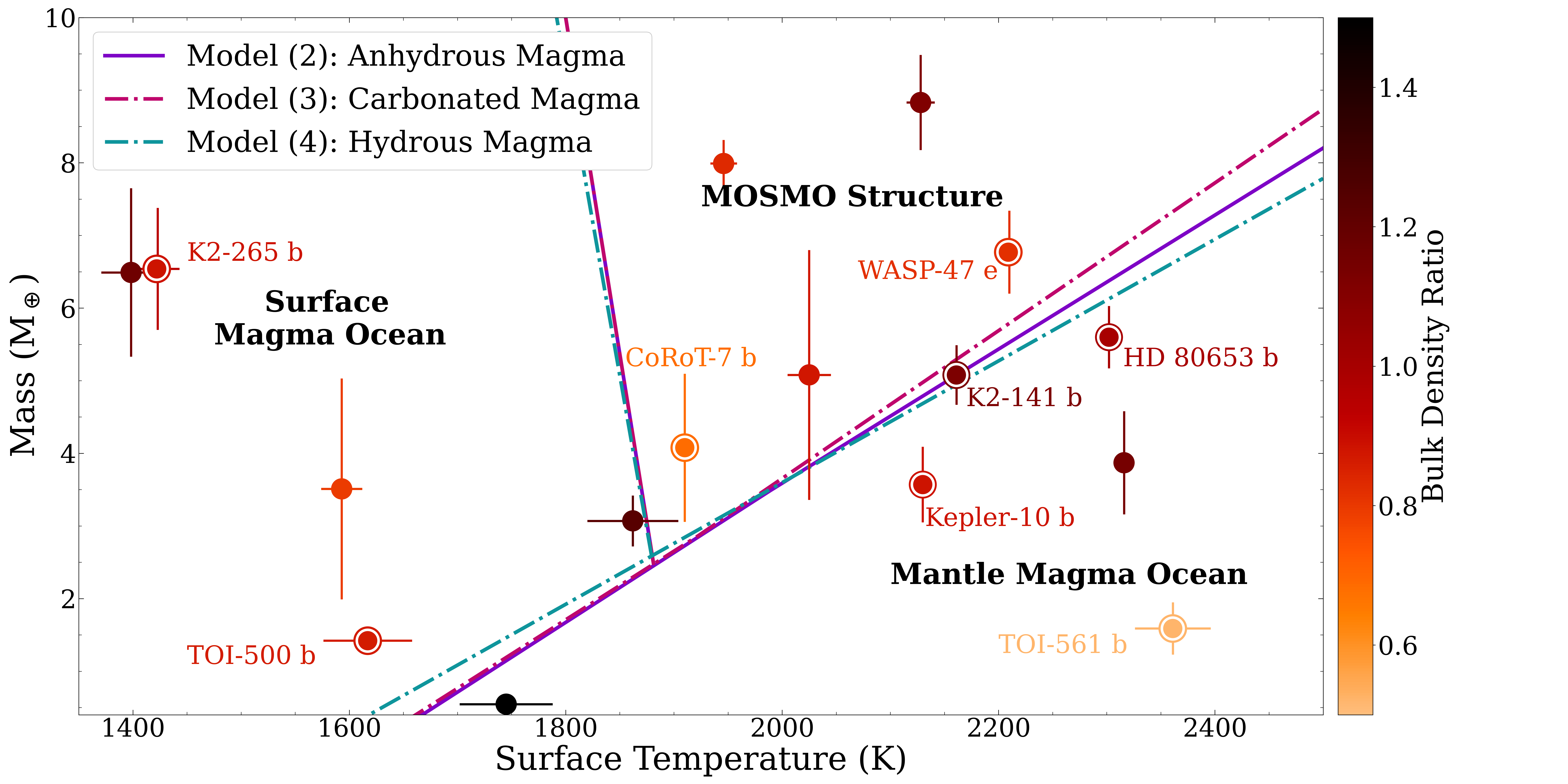}
\caption{{ Phase diagram for each model showing the mantle structure for a given mass and surface temperature. We include likely lava worlds selected using criteria described in \S \ref{s:MR}. The planets are colored by their bulk density ratio, which is their observed density divided by the expected density assuming an Earth-like composition. Planets distinguished with white circles and planet names are discussed in detail in section  \S \ref{s:lowest} and \S \ref{s:known}.}}
\label{fig:Phase}

\end{figure*}

\subsection{Interior Compositions}
To investigate the impacts of volatile compositions on the mass and radius of a planet, we consider four different model scenarios in which each scenario assumes a liquid iron core:
\begin{enumerate}
    \item  Solid rocky mantle
    \item  Solid rocky mantle and anhydrous melt
    \item  Solid rocky mantle and a carbonated melt
    \item  Solid rocky mantle and a hydrous melt
\end{enumerate}

Figure \ref{fig:Pslice} illustrates each model scenario with masses equal to four earth masses at a surface temperature of 2000 K demonstrating that for low masses generally, the hydrous magma composition has the largest radius while the solid planet has the smallest radius.

\section{Results} \label{s:MR}

Incorporating melting temperatures into our model, we find three distinct magma ocean mantle structures when surface temperatures exceed the solidus. Specifically, we find that the mantle of a lava world may be completely molten, only molten at the surface of the planet, or exhibit a layered magma ocean structure hereafter referred to as a MOSMO structure. Here, we define a mantle with a MOSMO structure as having a magma ocean  (MO) at the surface, a solid rock layer (S) mid-mantle, and a basal magma ocean (MO) at the base of the mantle (See Figure \ref{fig:Structure}). It is this intermediate solid rock layer that has the potential to serve as a mechanical barrier to volatile transport from a basal magma ocean to the surface. 

In Figure \ref{fig:Phase}, we show the “phase diagram” for the mantle structure including several known planets that are likely lava worlds. The planets are colored by their bulk density ratio, which is their observed density divided by the expected density assuming a solid, Earth-like composition with an Earth-like light element budget and core mass fraction. Bulk density ratios $<$ 1 indicate low-density planets whereas $>$ 1 indicate likely super-mercuries. We construct this planet sample using the Exoplanet archive making the following cuts. We require all planets within the sample to have mass and radius measurements. We remove any planets with flagged or controversial measurements. Using the period-dependent radius gap as an
upper bound on planet radius, we remove planets with larger radii \citep{VanEylen2018}.  We then explore the structure and bulk density of planets with surface temperatures above 1350 K assuming no atmosphere and surface albedo of zero. We choose 1350 K as it is the lowest temperature that is predicted by the anhydrous melt curve. Our results suggest that MOSMO structure and surface magma ocean planets are slightly more common than planets that are completely molten (see Figure \ref{fig:Phase}).


 We investigate the impact of surface temperature on the depth of the solid-melt transition. Using the median mass of our planet sample, 4 M$_\oplus$, we vary the surface temperature displaying the lower boundary of the surface magma ocean and the upper boundary of the basal magma ocean (see Figure \ref{fig:T-P}). As the surface temperature increases, the depth of the surface magma increases until the mantle no longer contains a solid layer at depth. We find that the planet becomes completely molten at a surface temperature ($T_{surf}$) of 2050 K  for Model (2) and (3) whereas Model (4) becomes mantle magma ocean at a $T_{surf}= 2100$ K. In the interval where a planet's mantle transitions from solid to mantle magma ocean, the planet will form a MOSMO structure.

\begin{figure}[!t]
\begin{center}
\includegraphics[width=\linewidth]{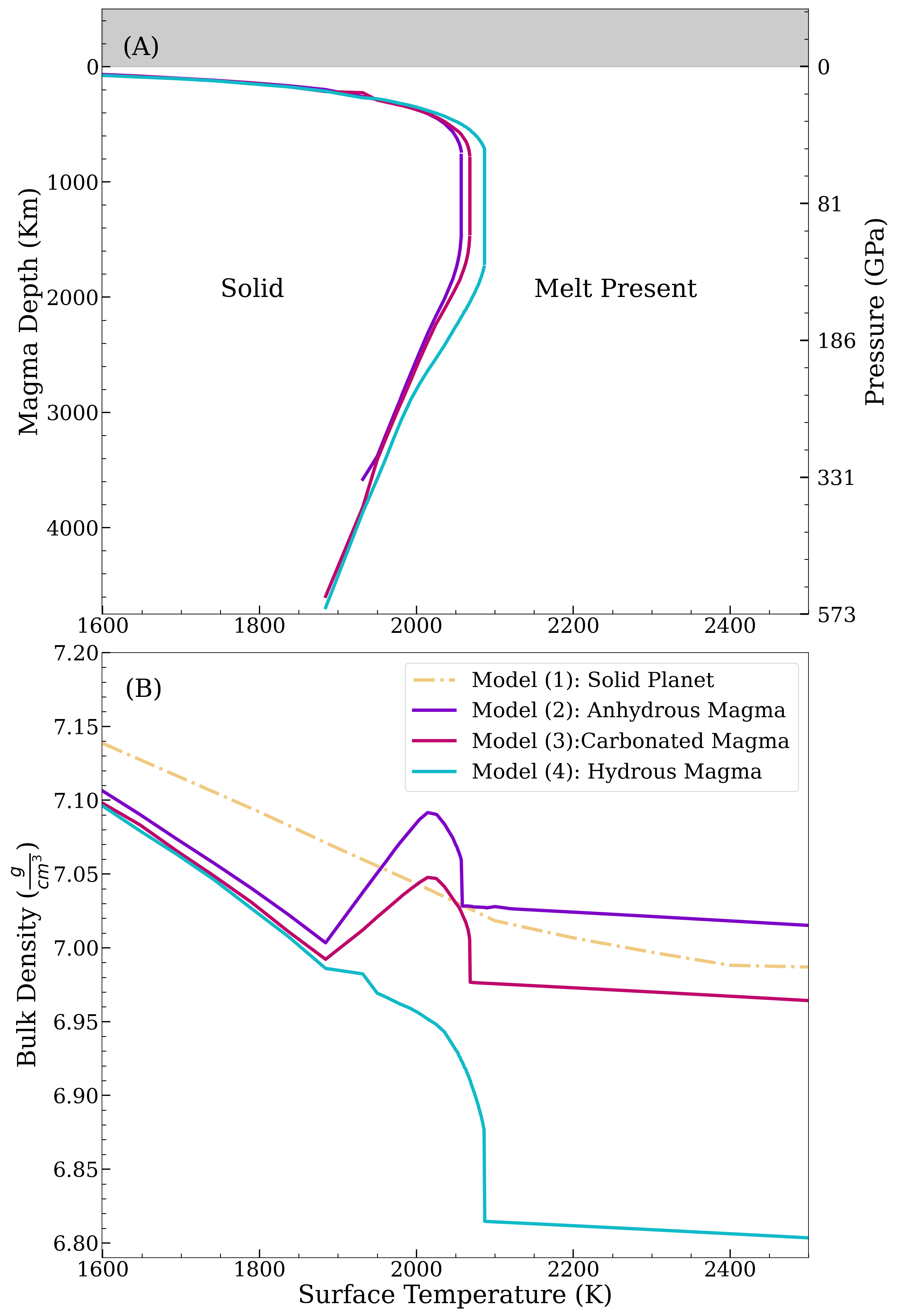}
\caption{(top) The depth of solid-melt transition as a function of surface temperature for a 4 M$_\oplus$ planet over a range of surface temperatures. The pressure on the y-axis corresponds to the hydrous melt (Model (4)). The core-mantle boundary pressure for all three scenarios are between 573-594 GPa. (bottom)  We show the bulk planet density for each model as a function of surface temperature. The increase in decrease for Models (2) and (3) reflect the competing effects of thermal expansion and compression for a planet with a MOSMO structure.  Additionally, we include Model (1) demonstrating the effects of thermal expansion on solid rock.}
\label{fig:T-P}
\end{center}
\end{figure}

Figure \ref{fig:T-P}{.A} shows that for a given mass, a planet will have a surface temperature range in which they exhibit a MOSMO structure that spans $T_{surf}\sim$ 2000 K for 4 M$_\oplus$ planets. The formation of a basal magma ocean occurs at $T_{surf} \sim$ 1872 K, but varies minimally depending on the magma volatile content. Any differences between the models are not significant given the assumptions of the melt curve at high pressures. Within this range, the majority of the melt is located within the basal magma with the surface melt only deepening by a few percent with increasing surface temperature. 

The depth of the surface magma ocean on a 4 M$_\oplus$ planet is dependent on the magma composition and the imposed surface temperature (See Figure \ref{fig:T-P}{.A}). Model (2), the anhydrous magma, has a surface magma ocean that reaches a depth of 68 km (4.8 GPa) given a surface temperature of 1600 K. Comparing Model (3) and Model (4) to Model (2), we find that Model (2) has a shallower surface magma for a given surface temperature with the most prominent depth difference occurring at $T_{surf}=1600$ K. At this surface temperature, the surface magma ocean of Model (3) and Model (4) have 10\% and 12\% greater depth than Model (2), respectively. The difference in surface magma depth between the models becomes less pronounced with increasing surface temperature. Before the mantle becomes mantle magma ocean for Model (2), the depth of the surface magma ocean reaches 732 km (28 GPa) at a $T_{surf} \sim 1900$ K. At 1900 K, Model (3) and (4) only reach $\sim$ 4 \% greater depth than Model (2). Overall, we find that the surface magma oceans for all model scenarios are shallow consisting of less than 5\% of the radius fraction before the mantle becomes mantle magma ocean.

The formation of a basal magma ocean on a 4 M$_\oplus$ planet is invariant with respect to the volatile composition of the mantle (See Figure \ref{fig:T-P}{.A}).   We find the basal magma ocean forms with a $T_{surf} \sim$ 1870 K), melting only at the base of the mantle, 570 GPa. As the surface temperature increases the top of the basal magma ocean rises to 118 GPa for a $T_{surf} \sim$ 2080 K. At this point the MOSMO structure transitions to a mantle magma ocean as the top and bottom of the solid, mid-mantle region meet. Therefore, the surface temperature at which  the basal magma ocean first forms relies on a significant extrapolation of the melting curve, the existence of the MOSMO structure only relies on solidus temperatures at pressures not requiring extrapolation. Uncertainties in the melting curve parameters in equation \ref{eq:Kaz}-\ref{eq:Simon} will shift the top and bottom of the solid mid-mantle region to higher or lower pressures, but will not eliminate the layer.

\begin{figure*}[!t]
\begin{center}
\includegraphics[width=\linewidth]{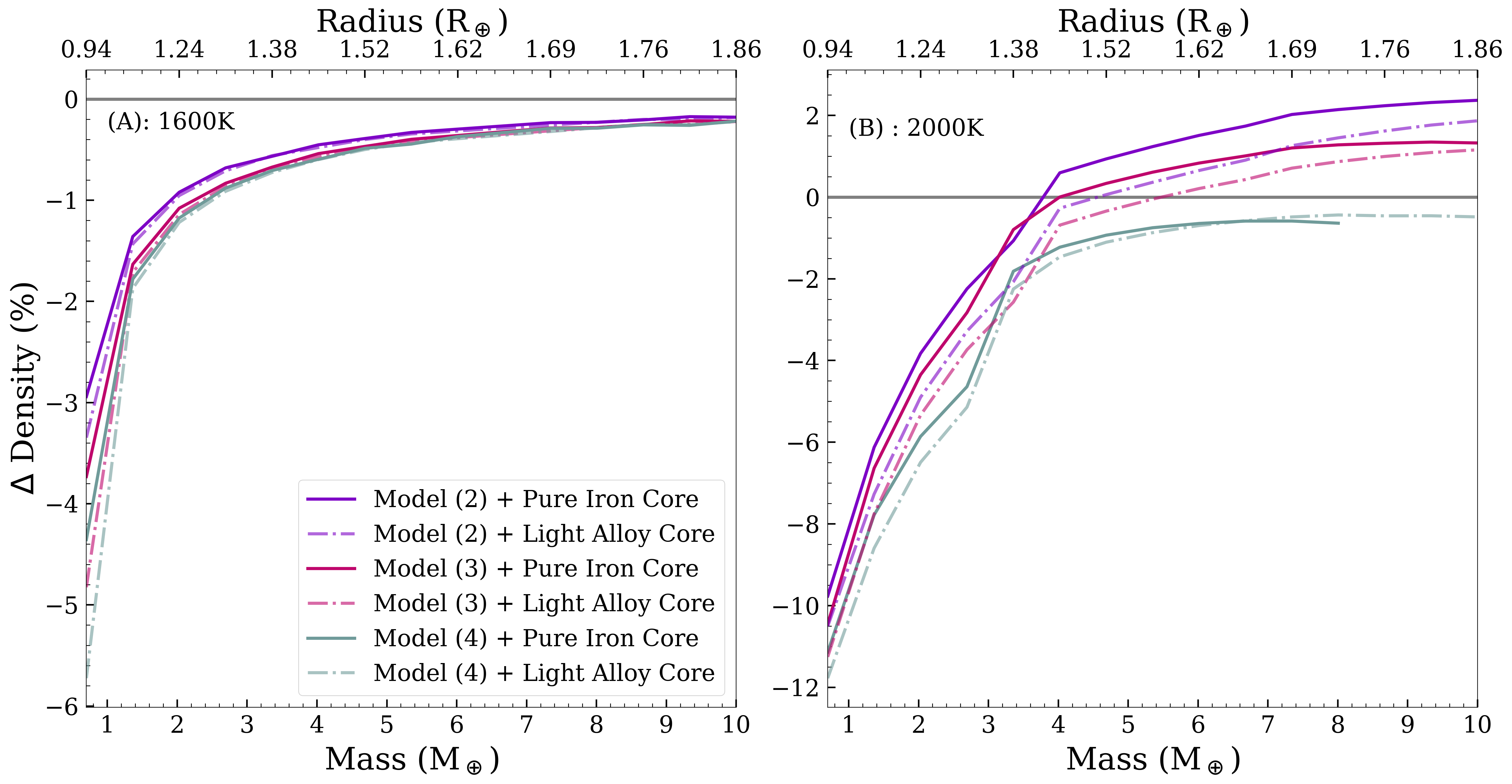}
\caption{The percent difference between the bulk density of each magma ocean planet and the solid planet with the corresponding core (i.e. light alloy core or pure iron core) at surface temperatures of (A) 1600 K and (B) 2000 K. The top axis shows radius values for  Model(2). The dashed lines represent models with light-element alloys in the core whereas solid lines show models with pure iron cores. Model (2), the anhydrous magma, is shown in purple. Model (3), the hydrous magma, is shown in teal and has the greatest decrease in density. Model (4), the carbonated magma, is shown in magenta. }
\label{fig:DensityDiff}
\end{center}
\end{figure*}

Although we show a 4 M$_\oplus$ planet in Figure \ref{fig:T-P}{.A}, we investigate the impact that mass has on magma depth as well. For planets $<$ 4 M$_\oplus$, we find that the MOSMO structure spans a smaller range of surface temperatures{, which can be seen in Figure \ref{fig:Density1}. The reduced surface temperature range results in a reduced mantle melt fraction of the basal magma ocean. For a 4 M$_\oplus$ at $T_{surf}1900$ K, the basal magma ocean would account for 19\% of the planet radius whereas a basal magma ocean on a 2.5 M$_\oplus$ at $T_{surf} 1900$ K would only account for 5.2 \% of the planet's radius.} The reduction in the basal magma ocean depth occurs to due the relative location of the solidus with respect to the geotherm}, the temperature-pressure gradient of the planet. The opposite is true for planets $>$ 4 M$_\oplus$ where MOSMO structure exists over a wider surface temperature range and at greater pressures.

We also include an analysis of the impact of surface temperature on the bulk density of a planet. Generally, we find that bulk density decreases with increasing surface temperature (Figure \ref{fig:T-P}.B).  The density decrease for Model (1) within Figure \ref{fig:T-P}.B demonstrates the effects of the thermal expansion of solid rock. The bulk density of a lava world decreases at a higher rate than that of a solid planet at surface temperatures that only produce a surface magma ocean demonstrating the greater thermal expansion of magmas at low pressures. However, Model (2) and Model (3) exhibit an increase in the bulk density within the surface temperature range where the MOSMO structure is present. This result indicates that the magma is highly compressed under the basal magma ocean pressure conditions, resulting in a density greater than solid rock under equivalent conditions. Therefore, the rapid increase in the basal magma fraction causes a local maximum in the bulk density before decreasing once the planet is above the solidus at all depths. Along with surface temperature, mass, core composition, and magma composition also impacts the bulk density of a planet.

\subsection{Impacts of Magma Composition} \label{s:magmacomposition}
To quantify the effects of magma and volatile composition on the density of the planet, we calculate the percent difference between the magma models and Model (1). Model (1), we use as the control model to determine the impact that the magma composition has on the mass and radius of the planet. Each subsequent model is compared with Model (1) shown in Figure \ref{fig:DensityDiff}. We set the mass to 4 M$_\oplus$ and surface temperature of each model to $T_{surf}=1600$ K and $T_{surf}=2000$ K considering the impact of surface temperature on differences in magma composition. 

Generally, we find that all four models behave similarly at $T_{surf}=1600$ K compared to $T_{surf}=2000$ K with any features being more prominent at higher surface temperatures. As planet mass increases, the density difference between a magma ocean and an equivalent mass solid planet decreases reflecting the effect of steeper pressure gradients in the planet and greater magma compressibility. At both 1600 K and 2000 K surface temperatures, Model (4) results in the greatest decrease in density whereas Model (2) results in the least. Broadly, all models find that low-mass planets exhibit the largest fractional inflation due to magma compared to higher mass planets. We will focus on the $T_{surf}=2000$ K results as it is the average surface temperature of the known low-density planets with surface temperatures above 1350 K. However, we briefly address the lower surface temperature model.  

As described above, we find that planets at $T_{surf}=1600$ K follow similar bulk density trends as a function of planet mass as those at $T_{surf}=2000$ K. Model (4) results in the lowest density being 4.3$\%$ less dense than an equivalent mass solid planet at a mass of 0.7 M$\oplus$. Model (2) results in the least density decrease at 2.9$\%$ at a mass of 0.7 M$\oplus$. As mass increases, we find that the decrease in the bulk density due to the presence of magma becomes less pronounced with the density difference being approximately 0.2\% for 10 M$_\oplus$. With the greatest density decrease for planets with a surface temperature of 1600 K being $<$ 4.3\%, the inflation due to magma would be indistinguishable from a solid planet with observational uncertainties on mass and radius. 

For planets at $T_{surf}=2000$ K, we find distinct differences from the $T_{surf}=1600$ K models. At 2000 K, Model (2) results in the least difference in density when compared to Model (1). It exhibits the most significant decrease in density of 9.8\% at lower masses and radii with the largest decrease at a mass and radius of 0.7 M$_\oplus$ and 0.93 R$_\oplus$, respectively. This reflects a planet with a mantle magma ocean mantle ocean with a core-mantle boundary pressure of 93.6 GPa. As mass and radius increase, we find that the anhydrous magma becomes more compressible.  Therefore, the anhydrous magma becomes denser than that of the solid planet (Model (1)). We find that this planet density crossover occurs at a mass and radius of 3.14 M$_\oplus$ and 1.45 R$_\oplus$. For the range of masses that we consider, we find that the highest density increase of 2.4\% compared to Model (1) at a radius of 1.83 R$_\oplus$.

The carbonated magma composition or Model (3) at $T_{surf}=2000$ K results in a moderate decrease in density of 10.4\% at a radius of 0.94 R$_\oplus$ and mass of 0.7 M$_\oplus$. Similar to Model (2), the magma becomes more compressible with increasing mass. At a radius of 1.51 R$_\oplus$ and 3.85 M$_\oplus$, Model (3) becomes denser than Model (1) with a maximum density increase of 1.3 \%. Compared to Model (2), the planet density crossover occurs at a higher mass and radius.

Model (4), the hydrous magma, results in the greatest decrease in density at 11.1\% compared to all other models at a radius of 0.94 R$_\oplus$, mass of 0.7 M$_\oplus$, and surface temperature of 2000 K. Given the EOS parameters for this magma composition (see Table \ref{tab:Params}), we expect the hydrous magma to start at a lower density. For the mass range that we consider, it would not become more compressible than the solid planet. However, we are limited to a smaller range of masses due to the hydrous model becoming dynamically unstable at masses greater than 8.7  M$_\oplus$.  The solid rock layer within the mantle becomes denser than the basal magma beneath requiring dynamical simulations.

\subsection{Light Elements in the Core} \label{s:LC}
We explore the impact of light elements in the core assuming an Earth-like composition. In Figure \ref{fig:DensityDiff}, we show the percent difference between the models with pure iron cores to the models with light alloys in the core.  As the mass increases, the difference in density of the magma compared to an equivalent mass solid planet is reduced as a result of the converging compressibility of the materials with increasing pressure (planet depth). When considering Earth-like compositions as described by Fe/Mg, the effect of light elements in the core is to reduce the planet's density due to the reduced mass of the core. However, the radius of the core increases to account for the added volume of oxygen when Fe/Mg remains constant. Assuming a pure iron core, a planet with a mantle magma ocean results in a bulk density difference of $\sim$ -9\% at 1 $M_\oplus$, and a MOSMO structure results in a $\sim$ 2\% density difference at 10 $M_\oplus$. The combined effects of a light alloy core and a magma ocean are $\sim$ -11\% and $\sim$ 1.5\%  at 1 $M_\oplus$ and 10 $M_\oplus$, respectively (Fig \ref{fig:DensityDiff}.B). Therefore, the impact on the bulk density is dominated by the effects of the magma ocean at constant Fe/Mg. This result suggests that the bulk density of an Earth-like composition lava planet with an Earth-like core light element budget and core mass fraction is more sensitive to the presence of magma than light elements in the core. The effects of the presence of light elements are magnified at greater core mass fractions potentially equivalent to the predicted density deficits of a surface magma ocean \citep{Unterborn2023}.

\newpage
\section{Discussion} \label{s:Discuss}

\subsection{Basal Magma Oceans}
\label{s:basal}
Several studies have proposed the potential for basal magma oceans to exist as a planet cools \citep{Pachhai2022,Labrosse2007,Elkins2005b}. Magma oceans may solidify in various ways: downward from the surface, upward from the core, or outward from the mid-mantle. The behavior of the solidification is a function of the mantle composition, and differences in gradient between the melting temperature and the local geotherm. When a planet solidifies from the mid-mantle, the magma ocean beneath the rock layer is a basal magma ocean. However, the inferred mechanical behavior of the intermediate rock layer changes given the melt curve that is used. 

We use the solidus melt curve to identify the existence of a solid rock layer separating the basal and surface magma. This differs from a liquidus melt curve as the rock layer would be a potentially porous crystal mush \citep{Marsh1989}. The resulting solid layer that we find could potentially trap volatiles deep within the mantle as there would not be a path for them to out-gas. Lava worlds that are continuously bombarded with radiation from their host star likely have a dry surface melt \citep{Kite2016, Leger2011}. Earth-sized planets have been shown to be inefficient at outgassing their volatiles \citep{SALVADOR2023,Miyazaki2022} over the timescales of the cooling Earth shortly after formation. However, it is unclear if long-lived magma oceans can retain their volatiles over long timescales. For this reason, the presence of a MOSMO structure is potentially significant for maintaining large volatiles within the mantle \citep{Bower2022,Caracas2019,Edmonds2018,Labrosse2007} even if the surface magma ocean eventually degasses.
{For a 4 $M_\oplus$ mass planet with a similar volatile concentration as Model (4) and (3), the presence of a basal magma ocean could potentially sequester approximately 130 times the mass of water than of Earth's present-day oceans and 1000 times the carbon in the Earth's surface and crust, respectively, for the water and carbon contents of the modeled magma.}

For simplicity, we do not consider a thermal boundary layer within the solid mid-mantle rock layer of the MOSMO structure. However, a non-convective solid rock layer with a thickness of a few kilometers would be enough to cause a thermal boundary layer \citep{ANDRAULT2019,monteux2016}, resulting in thinning of the mid-mantle solid layer. However, a fully convective system, in which material melts and solidifies as it crosses the solid material depth will reduce the net thermal boundary layer resulting from producing and consuming the heat of fusion.

\subsection{Lowest Density Planet Produced}\label{s:lowest}
\begin{figure}[!t]
\begin{center}
\includegraphics[width=\linewidth]{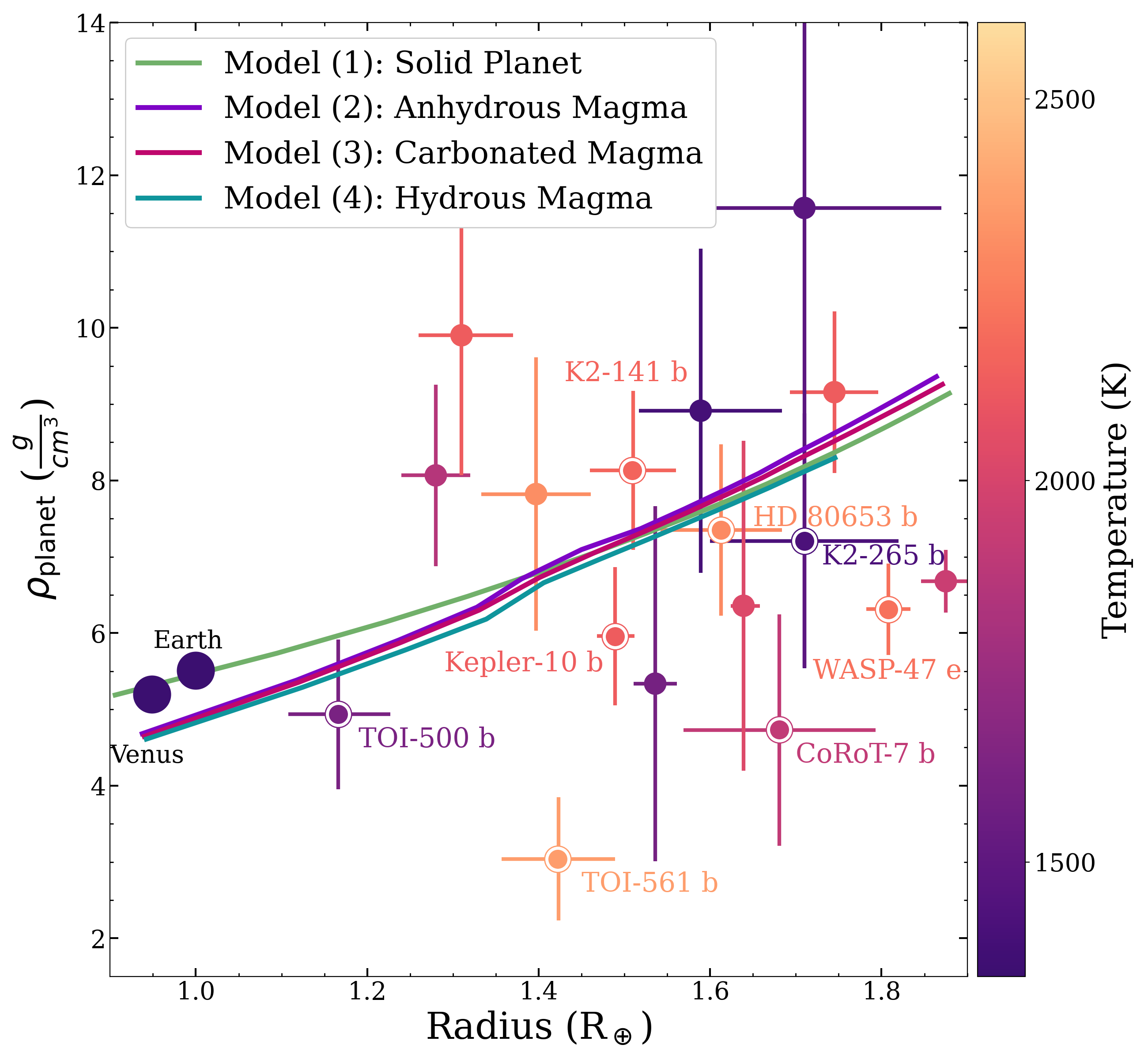}
\caption{Density-Radius plot for each model that is equivalent to a planet with a surface temperature of 2000 K.  We include likely lava worlds selected using criteria described in \S \ref{s:MR}. The planets are colored by their equilibrium temperatures. Planets distinguished with white circles and planet names are discussed in detail in \S \ref{s:lowest} and \S \ref{s:known}. Model (4), the hydrous magma, is limited to a radius of 1.7 R$_\oplus$, because the density of the magma becomes greater than the density of the solid in the surface magma ocean making it dynamically unstable.}
\label{fig:Density1}
\end{center}
\end{figure}

{To discuss the impact of magma on reducing the density of a planet, we consider our planet sample of likely lava worlds in terms of density and radius illustrated in Figure \ref{fig:Density1}. We include density-radius curves for each of our models for planets with surface temperatures of 2000 K. }Since magma is highly compressible, the greatest effects of its thermal expansion are seen at lower pressures or in Earth-sized planets as opposed to Super-Earths. Particularly, when considering larger Super-Earths masses (7-8 $M_\oplus$) our results suggest that magma ocean worlds without an atmosphere would be denser than their cooler counterparts. 

Within our planet sample, TOI-561 b, CoRoT-7 b  and TOI-500 b have the lowest bulk densities with densities of  3.00 $\pm$ 0.80 $\frac{g}{cm^3}$, $5 \pm 1.5$ $\frac{g}{cm^3}$ and $4.9 ^{+1.03}_{-0.88}$ $\frac{g}{cm^3}$, respectively \citep{Serrano2022,Lacedelli2021,Leger2011}. However, we find that the lowest bulk density produced by our models for the same planet mass is only consistent with the measurements for TOI-500 b. For TOI-561 b and CoRoT-7 b, our model overestimates the bulk density by 79\% and 50\%. If we instead calculate the density with light elements within the core discussed in \S \ref{s:LC}, we find a bulk density of $5.1$ $\frac{g}{cm^3}$ and $4.9$ $\frac{g}{cm^3}$ for TOI-561 b and CoRoT-7 b. These densities are still  69\% and 50\% greater. These results suggest that magma alone cannot account for the decrease in planets' bulk density, assuming they have Earth-like compositions. The impact of magma on the density of a planet is less than typical observational uncertainties on mass and radius, such that planets below the expected density for a rocky planet cannot be explained by the presence of melt.

Assuming an Earth-like composition for low-density planets requires the presence of an atmosphere regardless of the presence of a magma ocean. However, there is little evidence to suggest that a USP planet could sustain a substantial atmosphere that would significantly decrease the bulk density of a planet at such close proximity to its host star. Even assuming the presence of rock vapor atmosphere can only increase the radius of a super-Earth planet by approximately 1\%. For this reason, it may be more likely that these planets may have significantly different compositions from Earth with potentially lower core mass fractions \citep{Unterborn2023,Schulze2021}. 

\subsection{Known Lava Worlds}
\label{s:known}
75\% of the super-Earths that have been discovered to date are on orbits of less than 10 days. Given the presence of an atmosphere, $\sim$50\% of these planets have surface temperatures large enough to sustain a magma ocean \citep{Dorn2021}. However, without an atmosphere, the majority of these planets would still be able to maintain a magma ocean on their day-side.

Of the hundreds of super-Earths discovered, only 6 planets\footnote{\label{fnote}https://exoplanetarchive.ipac.caltech.edu, retrieved 27 Mar 2023} have been discovered with masses and radii that fall below the potential planet density crossover of 3.14 M$_\oplus$ and 1.45 R$_\oplus$ at surface temperatures greater than 1350 K. Therefore, the majority of the hot, relatively low-density planets exist in the regime where the presence of magma would cause them to have a slight increase in bulk density if they possess a magma ocean. For this reason, we cannot attribute the extremely low densities of some likely lava worlds primarily to magma. Instead, models addressing hot relatively low-density planets should consider an atmosphere or smaller core mass fraction in addition to magma.

To investigate the impact of magma on known planets, we select planets from our planet sample (Section \ref{s:MR}) that have a high probability of being atmosphere-free or having a thin atmosphere that could not result in significant changes in the density of the planet. When available, we select likely lava worlds from our sample evaluated within \cite{Schulze2021} that have a high probability that the inferred compositions from mass and radius alone are statistically indistinguishable from that of their host star at the $> 1\sigma$ level assuming no atmosphere. Therefore, we include K2-265 b and WASP-47e with high probabilities of having no more than a thin atmosphere and a core mass fraction comparable to their host star (94\% and 80\%, respectively) \citep{Schulze2021}. 

We also include three additional planets that are likely to be atmosphere-free or have thin atmospheres adopting the nominally rocky planet zone (NRPZ) introduced by \cite{Unterborn2023}. The NRPZ is the likelihood that the planet's observed mass and radius alone are consistent with the planet having a bulk rocky composition consistent without the addition of significant surface volatiles \cite{Unterborn2023}. We include HD 80653 b, Kepler-10 b, and K2-141 b with NRPZ probabilities of 66\%, 60\%, and 55\%, respectively. 

Below we discuss these five likely atmosphere-free lava worlds. Due to the current observational uncertainties of planet mass and radius for this population, we cannot distinguish between Models (1)-(4). However, we include a discussion on bulk density to demonstrate that magma alone cannot cause extremely low densities in planets. We also discuss the required observational uncertainties to distinguish between a solid planet and lava world for each planet. Therefore, we choose Model (4) as it produces the largest differences from Model (1), the solid planet. We also assume a pure iron core. 

We provide the following discussion primarily to consider the impacts that high surface temperatures may have on the structure of a planet. By relying on the underlying physics, we can gain insight into the potential structures that may arise assuming an Earth-like composition. We consider the potential melt fraction and structure of the mantle assuming an Earth-like composition.

\subsubsection{K2-265 b}
K2-265 b short-period planet discovered in 2018 on an orbit of 2.37 days with a density of $7.1\pm 1.8$ $\frac{g}{cm^3}$ \citep{Lam2018A}. It has a mass and radius of $6.54 \pm 0.84$ M$\oplus$ and  $1.71\pm 0.11$ R$\oplus$, respectively \citep{Lam2018A}.  Given the close proximity its host star, K2-265 b is exposed to strong stellar irradiation likely resulting in the photoevaporation and loss of its atmosphere \citep{Lopez2012}. Assuming the planet is a black body with a zero albedo, the day-side equilibrium temperature is calculated to be T$_{eq} \sim$ 1400 K.

Applying Model (4), we find that 1\% of the mantle is molten at the surface due the low equilibrium temperature. We find the bulk density to be 7.94 $\frac{g}{cm^3}$, 11\% greater than the measured density. The density is consistent with a lava world due to observation uncertainties; however, it is also indistinguishable from a solid planet as Model (1) also produces a bulk density of 7.95 $\frac{g}{cm^3}$. With such a small percentage of melt, there is minimal impact on the bulk density, and the surface magma ocean is likely entirely degassed. With the low equilibrium temperature, the uncertainties on mass and radius required to differentiate between a lava world with 1\% melt fraction must be less than 0.01\%.
 

{
\subsubsection{WASP-47 e}
WASP-47 e is an USP planet on an orbit of 0.7895 days\citep{Bryant2022}. It has an observed bulk density of $6.29 \pm 0.60$ $\frac{g}{cm^3}$ with a mass and radius of $6.77 \pm 0.57$ M$\oplus$ and  $1.808 \pm 0.026$ R$\oplus$ \citep{Bryant2022}. Assuming a zero albedo, the day-side equilibrium temperature of the planet is 2200 K. As a well-characterized super-Earth, previous studies conjecture that the low density of WASP-47 e may be due to a Ca and Al-rich interior or the presence of a magma ocean and secondary atmosphere \citep{Gupta2021, Dorn2019}. However, these studies do not account for effects due to magma within their models.  

Applying Model (4) to WASP-47 e, we find that the mantle has a MOSMO structure. The total melt within the mantle is 88\% by volume. The majority of the melt is within the basal magma ocean within 80\% whereas only 8\% of the melt within the mantle is in the surface magma ocean. The basal magma ocean, therefore, has the potential to trap a significant inventory deep in the interior over extended times. Dynamical models of transport across a solid intra-mantle would be required to determine if outgassing may still be feeding an atmosphere at the surface. 

Considering the density of WASP-47 e, Model (4) produces a density of 7.89 $\frac{g}{cm^3}$. Given the uncertainties on the mass and radius, Model (4) produces a radius that is above the uncertainties on the observed density. Model (1) produces similar results with the bulk density being 7.95 $\frac{g}{cm^3}$. Although Model (4) has a slightly lower density, the presence of a magma ocean would not increase the radius enough to account for the observed low density.  In order to distinguish between a solid or lava world for WASP-47 e, the uncertainties on mass and radius required are $\lesssim$ 0.14\% and 
$\lesssim$ 0.05\%, respectively. Therefore, invoking compositional variables or a thick, unobserved atmosphere is necessary to account for the low density, such as an atmosphere or a reduced core-mass fraction.

\subsubsection{HD 80653 b}

Discovered in 2020, HD 80653 b is an USP planet on an orbit of 0.719 days. It has a mass and radius $5.60\pm0.43$ M$\oplus$ of $1.613\pm0.071$ R$\oplus$ , respectively \citep{Frustagli2020}. The observed bulk density is $7.4 \pm 1.1$ $\frac{g}{cm^3}$. Using a zero albedo, the day-side equilibrium temperature of the planet is 2300 K. 

Using this Model (4), we find that the mantle of HD 80653 b is a mantle magma ocean structure where the magma ocean extends from the surface to the core-mantle boundary, and therefore unlikely to be able to trap volatiles in the interior over the age of the system. The calculated bulk density is 7.4 $\frac{g}{cm^3}$. With Model (1), we find the bulk density to be 7.7 $\frac{g}{cm^3}$. Therefore, the density of HD 80653 b is consistent with both a solid or lava world due to the uncertainties on the observed mass and radius. To differentiate between Model (1) and Model (4), the uncertainty required would be $\lesssim$ 1.8\% on the mass and $\lesssim$ 0.6\% on the radius of the planet.}

\subsubsection{Kepler-10 b}
As the first rocky planet discovered with Kepler, Kepler-10 b is a well-known USP planet on an orbit of 0.8374 days with a mass of $3.57^{-0.51}_{-0.53}$ M$\oplus$ and radius of $1.489^{-0.023}_{-0.021}$ R$\oplus$ \citep{Dai2019,Batalha2011}.  With an equilibrium temperature of T$_{eq}$ = 2130$^{120}_{-60}$ K, it is a low-density lava world that does not deviate far from a solid planet on the density-radius relationship. The observed bulk density is 6.0 $\pm$ 1.1 $\frac{g}{cm^3}$ \citep{Dai2019}. 

 We apply our models to Kepler-10 b with Model (4),  the mantle consists of a mantle magma ocean structure. Similar to HD 80653, a magma ocean structure that extends from the surface to core-mantle boundary is unlikely to be able to trap volatiles in the interior over the age of the system. The calculated bulk density is 6.6 $\frac{g}{cm^3}$. Applying Model (1), we find the bulk density to be 6.8 $\frac{g}{cm^3}$. Therefore, a solid or lava world is consistent with the observed density as they produce densities that are well within the observational uncertainties. The required uncertainties on mass and radius to distinguish between Model (1) and Model (4) are $\lesssim$  1.9\% and $\lesssim$  0.06\%, respectively.

\subsubsection{K2-141 b}
K2-141 b is another USP planet on an orbit of 0.2803 days with a density of 8.2$\pm$1.1 $\frac{g}{cm^3}$ \citep{Barragan2018, Malavolta2018}.  It has a mass and radius of $5.08\pm0.41$ M$\oplus$ and $1.51 \pm 0.05$ R$\oplus$ with an equilibrium temperature of T$_{eq}$ = 2161$^{120}_{-60}$ K \citep{Malavolta2018}. However, K2-141 is also an active star with strong magnetic activity \citep{Barragan2018}. Therefore, induction heating could also be a potential heat source for K2-141 b. There is evidence showing that K2-141 b is inconsistent with a thick atmosphere \citep{Malavolta2018,Zieba2022}. However, it could possess a thin rock vapor atmosphere, but this would not significantly lower the bulk density. For this reason, we include this planet assuming that is atmosphere-free as a thin rock vapor atmosphere would not significantly impact the density. 

Applying Model (4) to K2-141 b, we find the planet to have a MOSMO structure. The total melt fraction of the mantle is 82\% by volume with the basal magma ocean containing the majority of the melt ( 68\% by volume). Therefore, the surface melt accounts for only 14\%. As with WASP-47e, the MOSMO structure has the potential to trap volatiles in its interior with the basal magma ocean if the solid mid-mantle layer prevents mass transport to the surface.

Considering the bulk density of the K2-141 b, we apply Model (4). We calculate the bulk density to be 7.3 $\frac{g}{cm^3}$. Model (1) produces a bulk density of 7.4 $\frac{g}{cm^3}$. Due to the uncertainties on the mass and radius, both Model (1) and Model (4) are consistent with the observed density. To distinguish between the solid planet and magma ocean model, uncertainties on the mass and radius must be $\lesssim$ 0.2\% and $\lesssim$ 0.06\%, respectively. However, K2-141 b is a likely lava world that is denser than expected for a solid Earth-like planet. Therefore, Model (4) would be more appropriate, but the observational uncertainties required to distinguish between the models are not yet achievable. Given that it is denser than expected for an equivalent mass solid planet, K2-141 b may also have significant iron enrichment indicative of a Fe-rich super-Mercury in addition to a magma ocean.




\section{Summary \& Conclusions} \label{s:summary and conclusions}
{ Nearly half of the terrestrial planets discovered to date could maintain magma on their surfaces, particularly on their day-side. Therefore, it is important to understand the way in which magma oceans may impact the structures and observable properties of these likely lava worlds to better characterize them. Given that many of the likely lava worlds on USP have yet to detect substantial atmospheres large enough to significantly decrease its bulk density (e.g. \cite{Keles2022,Zieba2022,Kreidberg2019,Leger2011}), compositional factors such as the impact of magma or core mass fraction must be considered when characterizing these planets. }

In this paper, we investigated the impact of magma on the 1-D structure and bulk density of atmosphere-free magma ocean planets for the following magma compositions: anhydrous, hydrous, and carbonated magma. The objectives of this study were to determine whether a magma ocean is observable via the bulk density of a planet and to determine if volatiles may be trapped in the interior. Therefore, we constructed our model using a solidus melt curve placing an upper limit on the impact of magma on the bulk density of a lava world and placing a conservative lower limit on the mantle structure. From this study, we present our primary conclusions: 

\begin{enumerate}
    \item The presence of a magma ocean alone is not sufficient to explain low-density magma ocean planets that are expected to be atmosphere-free or have thin atmospheres (\S \ref{s:lowest}).
    \item For a given mass, there exists a range of surface temperatures in which a planet will have a basal magma ocean, which may sequester a significant amount of dissolved volatiles (\S \ref{s:MR}).
    \item The addition of H$_2$O or CO$_2$ to the magma does not significantly impact the calculated bulk density of a planet only resulting in a maximum density difference of $\sim$ 1 $\%$ (\S \ref{s:magmacomposition}).
    \item For magma ocean planets that are atmosphere-free, the presence of magma can impact the bulk density of the planet causing two distinct regimes where magma ocean planets exhibit a planet density crossover that is dependent on mass and surface temperature. This leads to two regions where a magma ocean planet may be more or less dense than an equivalent mass solid planet for anhydrous or carbonated magmas (\S \ref{s:magmacomposition}). 
    \item For an Earth-like core light element budget and core mass fraction, the addition of magma has a greater impact on the bulk density than the addition of lighter elements within the core for planets with masses and radii less than $\sim$3.14 M$_\oplus$ and $\sim$ 1.45 R$_\oplus$ (\S \ref{s:LC}).
    
\end{enumerate}

\acknowledgments
K.M.B. acknowledges support from the NSF Graduate
Research Fellowship Program under Grant No. (DGE-1343012). Any opinions,
findings, and conclusions or recommendations expressed in this material are those of the
author(s) and do not necessarily reflect the views of the National Science Foundation. WRP acknowledges support from NSF under Grant No. (EAR-1724693). CTU acknowledges support under grant NNX15AD53G. The results reported herein benefited from collaborations and/or information exchange within NASA's Nexus for Exoplanet System Science (NExSS) research coordination network sponsored by NASA's Science Mission Directorate. J.W. acknowledges the support by the National Science Foundation under Grant No. 2143400.

 \software{ \texttt{ExoPlex}~\citep{2018NatAs...2..297U,Unterborn2019, Unterborn2023}}

\bibliography{Bib}
\nocite{*}

\end{CJK*}
\end{document}